\documentclass[10pt,a4paper,twocolumn]{article}

\usepackage{graphicx} 

\usepackage{subfigure} 
\usepackage{subfigmat} 

\usepackage{dcolumn}           
\newcolumntype{.}   {D{.}{.}{-1}} 
\newcolumntype{d}[1]{D{.}{.}{#1}} 
\newcolumntype{e}   {D{E}{E}{-1}} 
\newcolumntype{E}[1]{D{E}{E}{#1}} 

\usepackage{amsmath}
\usepackage{mathptmx}

\usepackage{amssymb}
\usepackage{amsthm}

\usepackage{float}

\usepackage{setspace}

\usepackage{sectsty}

\newcommand{\myFontSize}{\fontsize{10}{12}\selectfont}
\renewcommand{\normalsize}{\fontsize{10}{12}\selectfont}
\sectionfont{\myFontSize}       
\subsectionfont{\rm\myFontSize\itshape} 

\usepackage{titlesec}
\titlespacing*{\section}{0pt}{10pt}{0pt}
\titlespacing*{\subsection}{0pt}{10pt}{0pt}

\usepackage{secdot}
\sectiondot{subsection}
\sectionpunct{section}{. }    
\sectionpunct{subsection}{. } 
\usepackage{tikz}
\usepackage{physics}

\newtheorem{remark}{Remark}

\usepackage{caption}
\usepackage{xcolor} 
\usepackage{fancyhdr} 
\usepackage{lastpage}
\renewcommand{\headrulewidth}{0pt} 

\usepackage[numbers]{natbib}

\begin{document}

%
\twocolumn[
\begin{@twocolumnfalse}

\vspace{0pt}
\begin{center}
    
    \vspace{15pt}
    \textbf{Probability of Collision of satellites and space debris for short-term encounters: Rederivation and fast-to-compute upper and lower bounds}
    
    \vspace{10pt}
    \textbf{
        \selectfont\fontsize{10}{0}\selectfont Ricardo~Ferreira~\textsuperscript{a*},~Cláudia~Soares~\textsuperscript{b},~Marta~Guimarães~\textsuperscript{c} 
    }
\end{center}


\vspace{-10pt} 
\begin{flushleft}
    \textsuperscript{a}\textit{
    \fontfamily{ptm}\selectfont\fontsize{10}{12}\selectfont FCT-UNL, Portugal}, \underline{rjn.ferreira@campus.fct.unl.pt}
    \\
    \textsuperscript{b}\textit{
        \fontfamily{ptm}\selectfont\fontsize{10}{12}\selectfont FCT-UNL, Portugal},
        \underline{claudia.soares@fct.unl.pt}
    \\
    \textsuperscript{c}\textit{
        \fontfamily{ptm}\selectfont\fontsize{10}{12}\selectfont Neuraspace, Portugal},
        \underline{marta.guimaraes@neuraspace.com}
    \\
    \textsuperscript{*}\fontfamily{ptm}\selectfont\fontsize{10}{12}\selectfont Corresponding Author  
\end{flushleft}


\begin{abstract}
The proliferation of space debris in LEO has become a major concern for the space
industry. With the growing interest in space exploration, the prediction of potential collisions between
objects in orbit has become a crucial issue. It is estimated that, in orbit, there are millions of fragments
a few millimeters in size and thousands of inoperative satellites and discarded rocket stages. Given the
high speeds that these fragments can reach, even fragments a few millimeters in size can cause fractures
in a satellite’s hull or put a serious crack in the window of a space shuttle.
The conventional method proposed by Akella and Alfriend in 2000 remains widely used to estimate
the probability of collision in short-term encounters. Given the small period of time, it is assumed that,
during the encounter: (1) trajectories are represented by straight lines with constant velocity; (2) there
is no velocity uncertainty and the position exhibits a stationary distribution throughout the encounter;
and (3) position uncertainties are independent and represented by Gaussian distributions.
This study introduces a novel derivation based on first principles that naturally allows for tight and fast
upper and lower bounds for the probability of collision. We tested implementations of both probability
and bound computations with the original and our formulation on a real CDM dataset used in ESA’s Collision Avoidance Challenge. Our approach reduces the calculation of
the probability to two one-dimensional integrals and has the potential to significantly reduce the processing
time compared to the traditional method, from 80\% to nearly real-time.


\end{abstract}

\end{@twocolumnfalse}

\twocolumn]


\section{Introduction}

Since the early days of space exploration, experts have warned that space debris could pose a long-term threat to the future of humankind in space. As pointed out by Donald Kessler \citep{kessler1978collision}, as the number of objects in orbit increases, the probability of collision also increases, resulting in a cascade effect that produces more fragments and, consequently, a greater number of objects in orbit. Nowadays, the Kessler effect is no longer a possibility and has become a real threat. Objects in orbit range from operating satellites to pieces of broken spacecrafts to tiny fragments like scraps of paint. In ESA's latest annual space environment report \citep{lemmens2020esa}, updated in December 2020, satellites in operation share their orbits with approximately 10,000 tons of space debris. 

It is estimated that, in orbit, there are millions of fragments a few millimeters in size, several thousand a few centimeters in size, to a few thousand inoperative satellites and discarded rocket stages. Of these numbers, only 26,000 objects are tracked from Earth. Smaller objects that we cannot monitor are estimated from statistical models \citep{lemmens2020esa}. This evolution leads to an increase in the attention of the scientific community to various problems, such as collision avoidance between satellites \citep{kleinig2022collision,bonnal2020just,mishne2017collision,reiland2021assessing} and collision risk \citep{le2018space,lucken2019collision,balch2019satellite}.

The process of tracking and monitoring objects in orbit is associated with uncertainty~\citep{poore2016covariance}, from which some types stand out: Structural uncertainty (or model bias); Sensor measurement noise; Propagation of uncertainty; and Algorithmic uncertainty. All these factors have a significant impact on conjunction assessment \citep*{poore2016covariance, klinkrad2006space}.

Given the high speeds that these fragments can reach, even fragments a few millimeters in size can cause fractures in a satellite's hull or put a serious crack in the window of a space shuttle \citep{pelton2013space}. Many of the satellites in orbit are small, as the Proba satellites, therefore such damage can become a critical threat. These assets play a crucial role for numerous services on the planet, such as communications, Earth observation, scientific experiments and weather monitoring. Also, the lifetime of these satellites is limited as they are dependent on the remaining fuel to adjust orbits or perform maneuvers to avoid collisions. Thus, it is of utmost importance that these maneuvers are only performed in situations of real danger to the integrity of these assets.

\paragraph*{Related Work}

Some methods were presented to address the risk of collision between objects in short and long-term encounters, with varied assumptions about the dynamical model. The short-term encounter model assumes that the encounter of the objects lasts for a very short period of time (in the order of seconds). Such approximation is typically used in Low-Earth Orbit (LEO), where the orbital velocities of the objects are high and, therefore, the relative motion can be assumed rectilinear. 

The evolution of the number of objects in LEO has been substantial in recent years compared to the evolution in other types of orbits \citep{lemmens2020esa}. This considerable evolution presents the short-term encounter model as central to the subject of conjunction assessment. Adopting this model for short-term encounters, many of the methods presented to compute the risk of collision only consider position uncertainty that remains stationary throughout the entire encounter and linear motion \citep*{akella2000probability,alfriend1999probability,chan1997collision,leclair2001probability,alfano2005numerical, patera2001general,foster1992parametric,berend1999estimation,klinkrad2006operational}, given the short period of time for the encounters.

The first methods presented for calculating the probability of collision in this model introduce the probability as an integral over a 2D circle centered at the origin \citep*{akella2000probability,foster1992parametric,klinkrad2006operational}, taking advantages of orthogonalities in the problem. Patera presents the collision probability as a one-dimensional path integral over the countour of the integration domain \citep{patera2001general} under strict assumptions. Alfano reduces the probability calculation to a one-dimensional integral using the error function \citep{alfano2005numerical} using approximations. Coppola takes into account uncertainty in velocity and abandons the assumption that the trajectories are linear. However, these methods imply computationally heavy procedures, where the probability of collision is given by a triple integral \citep*{coppola2012including,arzelier2020rigorous}.

Akella and Alfriend present a modification to the method by Foster \citep{foster1992parametric}, defining the probability of collision as the conditional probability of a collision occurring with respect that the minimum miss distance occurs at the estimated closest point of approach \citep{akella2000probability}. The simplification of the shape of objects as spheres and the simplicity of the method reducing the problem to an integral in two dimensions, made the method proposed by Akella and Alfriend well-known and is now widely used to compute the collision risk for objects in LEO. We focus on this method throughout this document, presenting a rigorous derivation from first principles and introducing fast-to-compute upper and lower bounds.

\paragraph*{Contributions}

With this work, we

\begin{itemize}
    \item Derive from first principles the expression for the probability of collision, under the same assumptions as \citep{akella2000probability} (Section~\ref{sec:poc-derivation});
    \item Introduce accurate and fast-to-compute upper and lower bounds of the probability of collision (Section~\ref{sec:poc-bounds});
    \item Study the computational complexity of our method when compared with the naive one (Section~\ref{sec:computational-complexity});
    \item Validate our approach on a real dataset of Conjunction Data Messages (CDM) (Section~\ref{sec:results}).
\end{itemize}

\section{Probability of Collision according to Akella and Alfriend}

Akella and Alfriend present an approach to compute the probability of collision between two objects in low-Earth orbit \citep{akella2000probability}. The authors define the probability of collision for the entire encounter to be the conditional probability of a collision occurring with respect that the minimum miss distance occurs at the estimated Closest Point of Approach (CPA). The miss distance is defined as the distance between the position vectors of both objects at the Time of Closest Approach (TCA). The objects are considered to have a spherical shape. As the duration of the encounters is very small, the following assumptions are made:

\begin{itemize}
    \item The trajectories are represented by straight lines with constant velocity, $x(t) = x_0 + t v$;
    \item There is no velocity uncertainty and the position exhibits a stationary distribution throughout the encounter;
    \item The position uncertainties are independent and represented by Gaussian distributions, $x_0 \sim \mathcal{N}\left(\bar{x}_0, \Sigma_{x_0} \right)$;
\end{itemize}

To clarify concepts, it is assumed that in a conjunction between two objects the designation of target for the asset to be protected in the conjunction (usually active satellites) and the designation of chaser for the object that can damage our target (mostly space debris). Based on these assumptions, the dynamical model for the two objects is defined as

\begin{equation}
    \label{eq:default-model}
    \begin{aligned}
        x(t) = x_0 + v_x t \quad \quad y(t) = y_0 + v_y t,
    \end{aligned}
\end{equation}
where $x_0$ and $v_x$ are the position and velocity vector of the chaser object, respectively, and $y_0$ and $v_y$ are the position and velocity vector of the target object, respectively.

With this it is possible to define the concept of range vector, $z(t)$, over time between the chaser and the target, i.e.,

\begin{equation}
    \label{eq:miss-vector}
    \begin{aligned}
        z(t) &= x(t) - y(t) = z_0 + v t,
    \end{aligned}
\end{equation}
where $z_0 = x_0 - y_0$ and $v = v_x - v_y$ are the relative position vector and relative velocity vector, respectively.

The authors proceed justifying that, based on geometry, the minimum distance satisfies the condition that the first derivative of the range vector norm squared with respect to time is equal to zero, to obtain the Time of Closest Approach (TCA), as

\begin{equation}
  \label{eq:tCPA}
  \begin{aligned}
    z_0^T v + v^T v t = 0 \Leftrightarrow TCA = -\frac{z_0^T v}{v^T v}.
  \end{aligned}
\end{equation}

The authors then define a new orthogonal system in terms of the velocity vectors of both objects and relative velocity. The new orthogonal system is defined by

\begin{equation}
  \label{eq:akella-orthogonal-system}
  \begin{aligned}
    \hat{i} = \frac{v}{\|v\|} \quad \quad \quad
    \hat{j} = \frac{v_x \times v_y}{\|v_x \times v_y\|} \quad \quad \quad
    \hat{k} = \hat{i} \times \hat{j}.
  \end{aligned}
\end{equation}
where $\alpha \times \beta$ is the cross product of vectors $\alpha$ and $\beta$. The plane $\hat{j}-\hat{k}$ is denoted as the conjunction plane. Since it is assumed that there is no velocity uncertainty and the $\hat{i}$ axis is collinear to the relative velocity vector, all the uncertainty is restricted to the conjunction plane. From unit vectors presented in~\eqref{eq:akella-orthogonal-system}, we can construct the $3 \times 3$ projection matrix, $C$, for the new orthogonal system, such that $CC^T$ = $C^T C$ = $I$. With this, we can obtain the $2 \times 6$ projection matrix, T, for the conjunction plane as

\begin{equation}
  \label{eq:akella-conjunction-projection}
  \begin{aligned}
    T^* = \begin{bmatrix}
            0 & 1 & 0 \\
            0 & 0 & 1
    \end{bmatrix} \quad \quad \quad
    T = T^* \begin{bmatrix}
            -C & C
    \end{bmatrix} 
  \end{aligned}
\end{equation}
Remembering that there is only uncertainty in the position vectors, taking into account the $3 \times 3$ covariance matrices, $\Sigma_{r_C}$ and $\Sigma_{r_T}$, of the chaser and target position vectors, respectively, the authors compute the covariance matrix relative to the range vector projected in the conjunction plane as

\begin{equation}
  \label{eq:akella-projected covariance}
  \begin{aligned}
    \Sigma^* = T \begin{bmatrix}
            \Sigma_{r_C} & 0 \\
            0 & \Sigma_{r_T}
    \end{bmatrix} T^T
  \end{aligned}
\end{equation}
where the covariance matrix, $\Sigma^*$, ends up being a $2 \times 2$ matrix. Finally, after the projection to the conjunction plane, together with the assumption that the uncertainties follow a Gaussian distribution, the authors define the collision probability as the integral over the circle centered on the origin of radius R, as

\begin{equation}
  \label{eq:akella-poc}
  \begin{aligned}
    P_C = \frac{1}{2 \pi |\Sigma^*|^{\frac{1}{2}}} \int\displaylimits_{-R}^{R} \int\displaylimits_{-\sqrt{R^2 - x_1^2}}^{\sqrt{R^2 - x_1^2}} \exp(S^*) \ \dd x_2 \dd x_1,
  \end{aligned}
\end{equation}
where 

\begin{equation}
  \label{eq:akella-poc-aux}
  \begin{aligned}
    S^{*} &= -\frac{1}{2} (x^{*} - z^{*})^T {\Sigma^{*}}^{-1} (x^{*} - z^{*}) \\
    x^* &= T^* C x, \quad \quad z^* = T^* C z_0,
  \end{aligned}
\end{equation}
and $R$ is the sum of the objects' radii. The determinant of matrix $A$ is denoted by $|A|$.
\section{Alternative derivation of the Probability of Collision}
\label{section:poc}

In this section, we detailed our derivation from first principles of the method for the computation of the probability of collision proposed by Akella and Alfriend \citep{akella2000probability}. Our derivation has two virtues: (1) we justify our final formulation with pure algebraic reasoning, and (2) it allows for an equivalent expression for the same measure, while opening the door to simple to compute lower and upper bounds, enabling large-scale selections of conjunctions.

\subsection{Problem Statement}

We start from the same dynamical model presented in \eqref{eq:default-model} with the same assumptions that there is no velocity uncertainty and the position uncertainties are constant throughout the encounter and are well-described by Gaussian distributions. Our goal is to calculate the \textit{miss distance}, $\|z(t_{CPA})\|$, which corresponds to the shortest distance between the two objects during the encounter. We can express this problem as an optimization problem as follows

\begin{equation}
    \label{pb:akella-minimization-problem}
    \begin{aligned}
        &\mbox{Minimize} \quad \left\|z(t)\right\|^2 \\
        &\mbox{subject to} \quad t \in \mathcal{T}
    \end{aligned}
\end{equation}
where $\mathcal{T}$ corresponds to the time interval of the encounter. Considering the initial time of identification of the possible conjunction as the origin, $\mathcal{T}$ corresponds to $\mathbb{R}_0^+$. Following this, we want to calculate the probability of collision, $P_C$, which corresponds to the probability of the \textit{miss distance} being less than the sum of the radii of the two objects ($R$),

\begin{equation}
    \label{eq:default-probability-of-collision}
    \begin{aligned}
        P_C = \mathbb{P}\left(\left\| z(t_{CPA}) \right\| \leq R \right) = \mathbb{P}\left(\left\| z(t_{CPA}) \right\|^2 \leq R^2 \right).
    \end{aligned}
\end{equation}

\subsection{Derivation of the Probability of Collision}
\label{sec:poc-derivation}

Starting from the definition of range vector presented in \eqref{eq:miss-vector}, we show that the minimum distance satisfies the condition that the first derivative of the range vector norm squared with respect to time is equal to zero.

The square of the norm of the $z(t) = z_0 + v t$ function is a convex function on $t$ \citep{boyd2004convex}. Therefore, Problem \eqref{pb:akella-minimization-problem} unfolds as a convex optimization problem. The shortest distance between the two objects occurs at the stationary point, i.e., where the first-order derivative is equal to zero, so

\begin{equation}
  \label{eq:tca}
  \begin{aligned}
    t_{CPA} = - \dfrac{v^T z_0}{v^T v}
  \end{aligned}
\end{equation}

\begin{equation}
  \label{eq:miss-distance}
  \begin{aligned}
    \|z(t_{CPA})\|^2 &= \left\| \left(I - \dfrac{v v^T}{v^T v}\right) \ z_0 \right\|^2 \\
    &= \left\| Q^T z_0 \right\|^2 . \qquad \mbox{Remark} \ \ref{remark:Q-projection-matrix}
  \end{aligned}
\end{equation}

\begin{remark}
\label{remark:Q-projection-matrix}
The term $I - \dfrac{v v^T}{v^T v}$ forms a projection onto the  orthogonal space of $v$. We see that the computation of the miss distance is the same as computing the norm of the projection of the range vector $z_0$ to the plane orthogonal to $v$ (a representation of what has been described is shown in Fig.~\ref{fig:error-vector}). Given the assumption that there is no uncertainty in velocity, then all uncertainty is contained in the two-dimensional plane orthogonal to the vector $v$ \citep{akella2000probability}. Therefore we resort to the $3 \times 2$ projection matrix Q, which projects to the plane orthogonal to $v$. We emphasize that Q is not a $3 \times 3$ rotation. Instead, Q is a rotation and projection. Considering the first element, $v_1$ of the relative velocity vector $v$, with $v_1$ being non-negative, and $v_*$ as the vector containing the remaining elements of $v$, i.e., $v = \begin{bmatrix} v_1 & v_*\end{bmatrix}$, we can compute $Q$ as \citep{horn2012matrix}

\begin{equation}
    Q = \begin{bmatrix}
    v_*^T \\
    \\
    -I + \frac{1}{1+v_1} v_* v_*^T
    \end{bmatrix}.
\end{equation}
\end{remark}
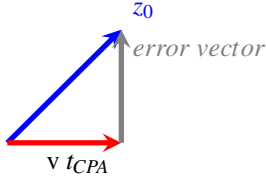
\begin{figure}
    \centering
    \begin{tikzpicture}
    \draw[line width=2pt,blue,-stealth](0,0)--(1.5,1.5) node[anchor=south west]{$z_0$};
    \draw[line width=2pt,red,-stealth,text=black](0,0)--(1.5,0) node[anchor=north east]{v $t_{CPA}$};
    \draw[line width=2pt, gray,-stealth](1.5,0)--(1.5,1.5) node[anchor=north west]{$\boldsymbol{error \ vector}$};
    \end{tikzpicture}
    \caption{Representation of the error vector between vector $z_0$ and vector $v$ times $t_{CPA}$.}
    \label{fig:error-vector}
\end{figure}
With this, joining the initial definition for $P_C$ in \eqref{eq:default-probability-of-collision} and the result obtained in \eqref{eq:miss-distance}, we reduce the probability space to two dimensions obtaining

\begin{equation}
  \label{eq:poc-transformed}
  \begin{aligned}
    P_C = \mathbb{P}(\| Q^T z_0 \|^2 \leq R^2).
  \end{aligned}
\end{equation}
In order to determine the probability distribution parameters described by $z_0$, we calculate the expected value and the variance. Given that $z_0 = x_0 - y_0$ the average value of $z_0$ is equal to the difference of the average values of $x_0$ and $y_0$

\begin{equation}
  \label{eq:average-value}
  \begin{aligned}
    \mathbb{E}(z_0) = \bar{x}_0 - \bar{y}_0 = \bar{z}_0,
  \end{aligned}
\end{equation}
where $\bar{x}_0$ and $\bar{y}_0$ are the mean values of the chaser and the target positions, respectively. Given the assumption of independence between the uncertainties in the positions, the covariance is equal to the sum of the covariances of $x_0$ and $y_0$

\begin{equation}
  \label{eq:cov-value}
  \begin{aligned}
    \mbox{cov}(z_0) = \Sigma_{x_0} + \Sigma_{y_0} = \Sigma,
  \end{aligned}
\end{equation}
where $\Sigma_{x_0}$ and $\Sigma_{y_0}$ are the $3 \times 3$ covariance matrices corresponding to the position uncertainty of the chaser and the target, respectively. Thus, we come to the conclusion that

\begin{equation}
  \label{eq:a-distribution}
  \begin{aligned}
    z_0 &\sim \mathcal{N}(\bar{z}_0,\,\Sigma) \\
    Q^T z_0 &\sim \mathcal{N}(Q^T \bar{z}_0,\,Q^T \Sigma Q).
  \end{aligned}
\end{equation}
To ease notation, we define the following variables: $w = Q^T z_0$, $\mu = Q^T \bar{z}_0$ and $\Sigma^* = Q^T \Sigma Q$. We can do an eigenvalue decomposition where $\Sigma^* = U \ \Lambda \ U^T$, such that $U$ is the $2 \times 2$ orthonormal eigenvector matrix and $\Lambda$ is a $2 \times 2$ diagonal matrix containing the eigenvalues for each axis. By applying the matrix $U$ to the random variable $w$, we have

\begin{equation}
  \label{eq:eigen-expected-value}
  \begin{aligned}
    \mathbb{E}(U^T w) &= U^T \mu \\
    &= \overline{w}
  \end{aligned}
\end{equation}
\begin{equation}
  \label{eq:eigen-variance}
  \begin{aligned}
    \mbox{cov}(U^T w) &= U^T \Sigma^* U \\
    &= U^T U \Lambda U^T U \\
    &= \Lambda
  \end{aligned}
\end{equation}
\begin{equation}
  \label{eq:eigen-distribution}
  \begin{aligned}
    U^Tw \sim \mathcal{N}(\overline{w},\,\Lambda).
  \end{aligned}
\end{equation}
Taking into account the expression in (\ref{eq:poc-transformed}), we obtain

\begin{equation}
  \label{eq:eigen-poc-formulation}
  \begin{aligned}
    P_C = \mathbb{P}(\| Q^T z_0 \|^2 \leq R^2) = \mathbb{P}(\| U^Tw \|^2 \leq R^2)
  \end{aligned}
\end{equation}
By taking the integral over the disk centered on the origin of radius R, we can compute the probability of collision as

\begin{equation}
  \label{eq:final-poc}
  \begin{aligned}
    P_C &= \dfrac{1}{2\pi |\Lambda|^{\frac{1}{2}}} \iint_{B_{0,R}} \exp(-\dfrac{1}{2} (x - \overline{w})^T \Lambda^{-1} (x - \overline{w})) \ \dd x \\
    &= \dfrac{1}{2\pi \sqrt{\lambda_1 \lambda_2}} \iint_{B_{0,R}} \ F_1(x_1) F_2(x_2) \ \dd x_2 \dd x_1,
  \end{aligned}
\end{equation}
where 

\begin{equation}
  \label{eq:final-poc-aux-functions}
  \begin{aligned}
    F_i(x) = \exp(-\frac{(x - \overline{w}_i)^2}{2 \lambda_i}), \quad i = 1,2,
  \end{aligned}
\end{equation}
$\lambda_i = \Lambda_{ii}$ is the $i$-th eigenvalue, and $x_i$ and $\bar{w}_i$ corresponds to the $i$-th component of $x$ and $\bar{w}$, respectively. This formulation simplifies the computation of the upper and lower bounds of the probability of collision, as we will show next.

\section{Upper and Lower Bounds}
\label{sec:poc-bounds}

\begin{figure}[h!]
    \centering
    \includegraphics[scale=0.5]{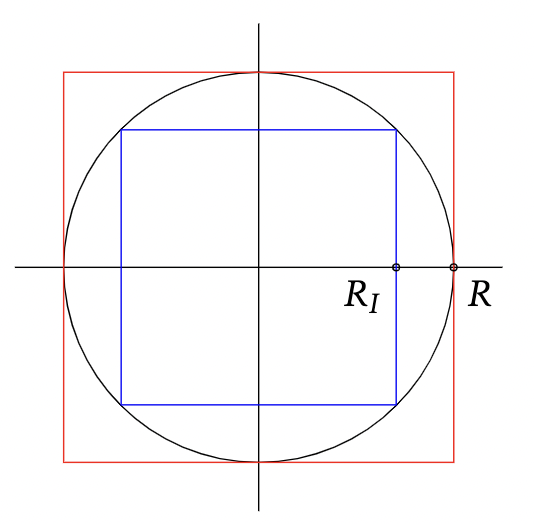}
    \caption[Graphical representation of the integration shapes for the $P_C$ and its upper and lower bound.]{Graphical representation of the 2D ball centered at the origin with radius $R$, and the red and blue rectangle represent the upper-bound and lower-bound of the $P_C$, respectively. The red rectangle has side $2R$ and the blue rectangle has side $2R_I$ such that $R_I = R \cos(\frac{\pi}{4})$.}
    \label{fig:upper-lower-representation}
\end{figure}
Viewing a graphical representation of the ball centered on the origin of radius $R$, as depicted in Fig.~\ref{fig:upper-lower-representation}, it is easy to see that it is possible to calculate an upper-bound and a lower-bound for the collision probability, as the $P_C$ will always be less than the integral over the red square and greater than the integral over the blue square. Thus, we obtain the expressions for the upper-bound and lower-bound of the $P_C$,

\begin{equation}
  \label{eq:upper-bound}
  \begin{aligned}
    \overline{P_C} &= \dfrac{1}{2\pi \sqrt{\lambda_1 \lambda_2}} \int\displaylimits_{-R}^{R} \int\displaylimits_{-R}^{R} F_1(x_1) F_2(x_2) \ \dd x_2 \dd x_1 \\
    &= \dfrac{1}{\sqrt{2\pi \lambda_1}} \int\displaylimits_{-R}^{R} \ F_1(x_1) \ \dd x_1 \ \dfrac{1}{\sqrt{2\pi \lambda_2}} \int\displaylimits_{-R}^{R} F_2(x_2) \ \dd x_2
  \end{aligned}
\end{equation}

\begin{equation}
  \label{eq:lower-bound}
  \begin{aligned}
    \underline{P_C} &= \dfrac{1}{2\pi \sqrt{\lambda_1 \lambda_2}} \int\displaylimits_{-R_I}^{R_I} \int\displaylimits_{-R_I}^{R_I} F_1(x_1) F_2(x_2) \ \dd x_2 \dd x_1 \\
    &= \dfrac{1}{\sqrt{2\pi \lambda_1}} \int\displaylimits_{-R_I}^{R_I} \ F_1(x_1) \ \dd x_1 \ \dfrac{1}{\sqrt{2\pi \lambda_2}} \int\displaylimits_{-R_I}^{R_I} F_2(x_2) \ \dd x_2.
  \end{aligned}
\end{equation}
The expressions for the upper and lower bounds can also be written using the error function as

\begin{equation}
  \label{eq:derivation-upper-bound-erf}
  \begin{aligned}
    \overline{P_C} &= G(R, \overline{w}_1, \lambda_1) \ G(R, \overline{w}_2, \lambda_2)
  \end{aligned}
\end{equation}

\begin{equation}
  \label{eq:derivation-lower-bound-erf}
  \begin{aligned}
    \underline{P_C} &= G(R_I, \overline{w}_1, \lambda_1) \ G(R_I, \overline{w}_2, \lambda_2),
  \end{aligned}
\end{equation}
where

\begin{equation}
  \label{eq:erf-aux-function}
  \begin{aligned}
    G(R, w, \lambda) = \frac{1}{2} \left[ \erf \left(\frac{R - w}{\sqrt{2 \lambda}} \right) - \erf \left(\frac{-R - w}{\sqrt{2 \lambda}} \right) \right],
  \end{aligned}
\end{equation}
with the error function defined as

\begin{equation}
  \label{eq:erf-definition}
  \begin{aligned}
    \erf(z) = \frac{2}{\sqrt{\pi}} \int_{0}^{z} \ \exp(-t^2) \dd t
  \end{aligned}
\end{equation}
It is a faster way to compute the upper and lower bounds since the error function approximations in use consist of a composition of arithmetic operations (such as additions and multiplications).

This derivation presents in detail all the steps from first principles to the final expression of the probability of collision. With this derivation, we introduce easy-to-compute upper and lower bounds for the probability of collision, which reduce to the product of two integrals in $\mathbb{R}$, which are easier to compute and can also be formulated using the error function.

\section{Computational complexity}
\label{sec:computational-complexity}

The claim that the presented solution is easier to compute can be mathematically justified. The base method integrates the function in two dimensions and therefore resorts to computationally heavier solutions. Dividing the domain into $M^2$ small squares implies an evaluation of the integrand function $M^2$ times, $O(M^2)$, and using Monte Carlo integration, we have a complexity of $O(N)$ such that $N$ is the number of samples for which we evaluate the integrand function and the error is proportional to $\frac{1}{\sqrt{N}}$. We could also use Gaussian Processes which implies evaluating the integrand function for $N$ samples, however, it presents a complexity of $O(N^3)$ due to the inversion of the kernel matrix in the Gaussian Process.

Our bounds reduce to the product of two one-dimensional integrals, which are easier to compute. We can resort to Simpson's rule whose composite form has a complexity of $O(N)$ such that the error is proportional to $\frac{1}{N^4}$, where $N$ is the number of subintervals for which we evaluate the integrand function. Computing the integral by evaluating the error function, $\erf(x)$, twice to obtain the probability value, since this function consists of a set of arithmetic operations (such as additions and multiplications), we have a computational complexity of $O(1)$.

\section{Data Exploration and Numerical Results}
\label{sec:results}

To test the presented solution, we apply the method proposed by Akella and Alfriend as well as our proposed method on a Conjunction Data Message (CDM) dataset used in ESA's Collision Avoidance Challenge \citep{uriot2021spacecraft}. More details about the dataset can be found on the challenge's page \footnote{https://kelvins.esa.int/collision-avoidance-challenge/data/}.

Not all features are necessary for the purpose of this paper. After analysis, we reduced the dataset to just 42 features, that describe the position and velocity of objects, their dimensions and their respective covariances, probability of collision and other factors for orbit determination such as eccentricity, inclination and the semi-major axis. We removed all dataset samples that contained at least one of the features without any recorded value. 

The highest degree of uncertainty is associated with the transverse (or along-track) component. This happens due to the impact of non-conservative and perturbative forces on the motion of the object and also due to the high velocities of objects in LEO, which lead to a great uncertainty when measuring the object's position with sensors on the ground \citep{metzthesis,vallado2001fundamentals}. In order to remove some outliers and unrealistic cases like those where the transverse component standard deviation is greater than the Earth's radius, we decided to remove samples beyond the $95$-th percentile of the chaser's transverse component standard deviation, given that the chaser reveals a greater uncertainty associated with the position. We are left with a dataset with 131,077 samples, which corresponds to 9,194 distinct events, and 42 features.

\paragraph*{Issues regarding $t_0$}

It is important to note that the position and velocity vectors, as well as the associated covariances present in the CDM, do not correspond to their values at the date of issue of the CDM, but are values propagated to the Time of Closest Approach (TCA). This fact strengthens our confidence on the linear motion assumption.

To assess the validity of our rederivation and evaluate the computational advantage of our method, we tested implementations of both probability and bound computations with the original and our formulation on the dataset described earlier. To evaluate our solution, we use the relative risk error metric between the outputs of both methods

\begin{equation}
    \varepsilon = \frac{\log_{10}(\hat{P}_C) - \log_{10}(P_{C_{AA}})}{\log_{10}(P_{C_{AA}})} \times 100,
\end{equation}
where $P_{C_{AA}}$ denotes the probability of collision obtained with the expression presented by Akella and Alfriend (AA), as a gold standard in use in industry, and $\hat{P}_C$ denotes the predicted value of our method.

\begin{figure}[ht!]
    \centering
    \includegraphics[scale=0.5]{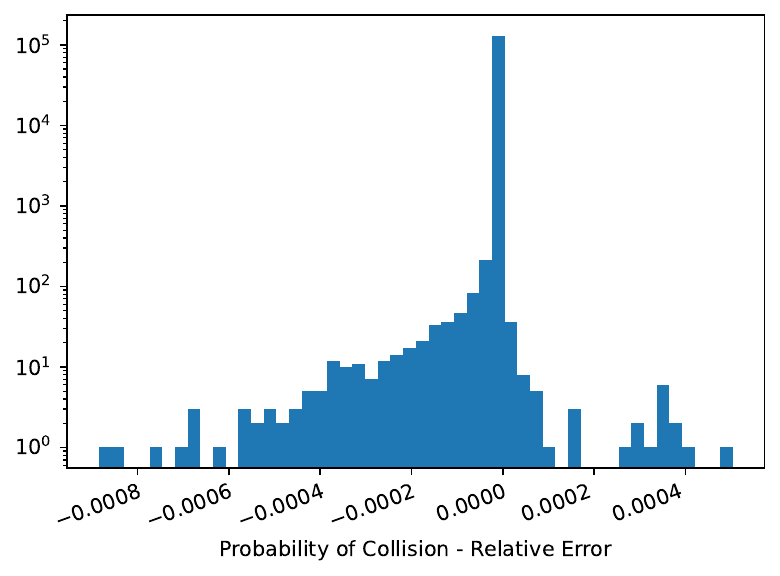}
    \caption[Relative error of the probability of collision computed by the proposed method]{Relative error of the probability of collision computed by the proposed method when compared to the base method, obtaining a maximum relative error in absolute value of $0.0008\%$.}
    \label{fig:poc_relative_diff}
\end{figure}

As we can see in Fig.~\ref{fig:poc_relative_diff}, our proposal obtains similar results compared to the method proposed by Akella and Alfriend \citep{akella2000probability}. The relative error between the probability values is concentrated around zero and the maximum relative error in absolute value is $8 \times 10^{-4}\%$. After analyzing the cases with the larger relative error, we found that these are rounding errors. Regarding processing times, as expected, both methods have similar processing times: the method proposed by Akella and Alfriend has an average processing time of $20.3$ ms and our rederivation has an average processing time of $11.8$ ms.

\begin{figure}[ht!]
    \centering
    \includegraphics[scale=0.5]{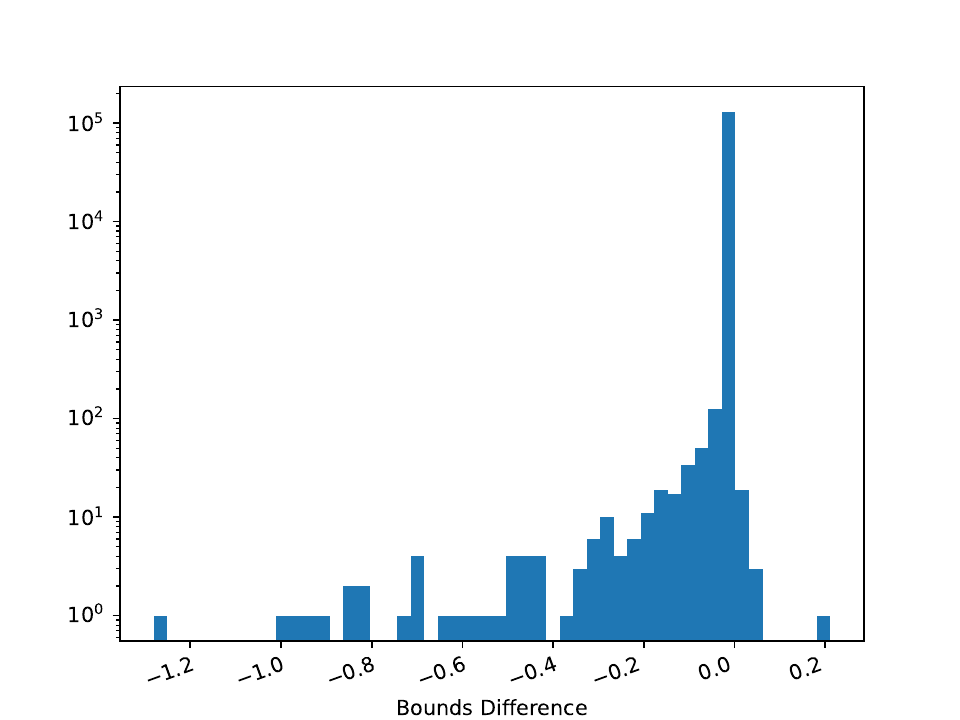}
    \caption[Difference between upper bounds computed by the proposed method and the naive one.]{Difference between upper bounds computed by the proposed method and the naive one. It can be seen that the proposed method presents in most cases smaller values and therefore closer to the real probability value, serving as a tighter bound of the probability of collision.}
    \label{fig:bounds_diff}
\end{figure}

Since the upper bound is more interesting than the lower bound, giving information of a slightly higher but safer value of the probability of collision, we compare the upper bound of our solution with the upper bound of the base method obtained by integrating the expression \eqref{eq:akella-poc} over the square of side $2R$, as depicted in Fig.~\ref{fig:upper-lower-representation}, i.e.,

\begin{equation}
  \label{eq:akella-bounds}
  \begin{aligned}
    \overline{P_C}_{AA} = \frac{1}{2 \pi |\Sigma^*|^{\frac{1}{2}}} \int\displaylimits_{-R}^{R} \int\displaylimits_{-R}^{R} \exp(S^*) \ \dd x_2 \dd x_1 .
  \end{aligned}
\end{equation}
To evaluate the accuracy of our bounds, we analyze the difference between the upper bounds of both implementations as

\begin{equation}
    \mbox{Bounds Difference} = \log_{10}(\overline{P_C}) - \log_{10}(\overline{P_C}_{AA}).
\end{equation}
The results obtained with this metric, as can be seen in Fig.~\ref{fig:bounds_diff}, shows that the difference is mostly zero or less, which means our upper bound implementation has higher accuracy, which can be useful to get a safe probability of collision with the advantage of being computationally faster.

\begin{figure}[ht!]
    \centering
    \includegraphics[scale=0.45]{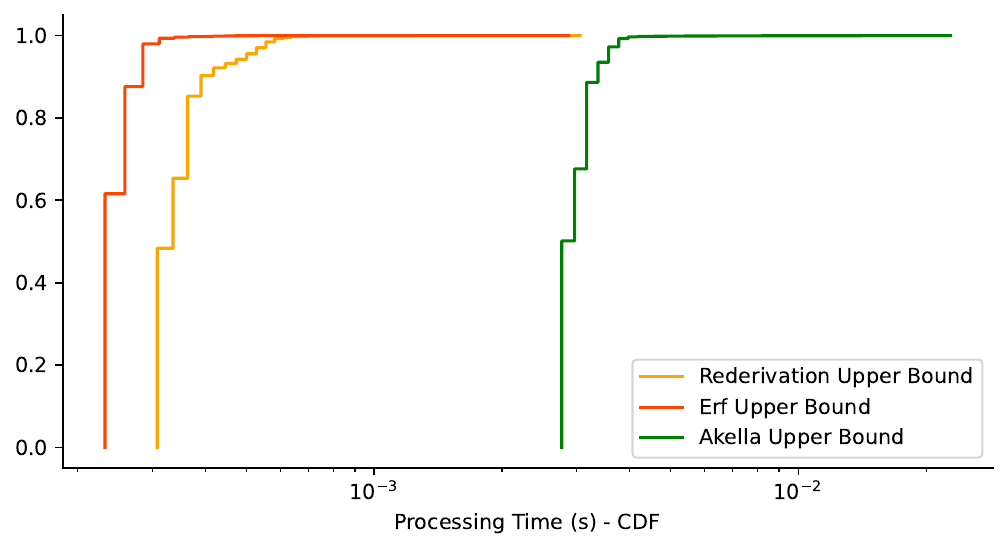}
    \caption{Cumulative distribution function of the processing times (in seconds) of the upper bound for the various implementations. It is possible to observe that the proposed method allows a faster processing time than the naive one, with the implementation using the error function being a little faster.}
    \label{fig:time_cdf}
\end{figure}

Although the complexity analysis shows our method can be faster, we assess the gain in processing time experimentally by recording the processing time of the upper bounds using~\eqref{eq:akella-bounds} for the upper bound with the approach based on Akella and Alfriend, ~\eqref{eq:upper-bound} for the upper bound using our rederivation and ~\eqref{eq:derivation-upper-bound-erf} using our rederivation with resort to the error function. As can be seen in Fig.~\ref{fig:time_cdf}, the processing time of our implementation in a standard laptop is noticeably faster than the processing time of the upper bound with Akella and Alfriend expression. While the upper bound with Akella and Alfriend expression has an average processing time of $3.2$ ms, our implementation using~\eqref{eq:upper-bound} has an average processing time of $0.32$ ms, which means that our solution is approximately $10 $ times faster. When using the implementation with resort to the error function, we obtain an average processing time of $0.24$ ms, which is approximately $13$ times faster.


However, when compared to the processing time to obtain the probability of collision, the gain is even more remarkable. Comparing with the average processing time of $20.3$ ms to calculate the probability of collision using the Akella and Alfriend approach we get a gain of $63.4$ times faster and $84.6$ times faster using the implementation with the integration feature and using the implementation using the error function, respectively.

Since the upper bound obtained through our rederivation presents values closer to the true probability value, as shown above in Fig.~\ref{fig:bounds_diff}, this solution presents a significant contribution to conjunction assessment. It is important to point out that the implementation using the error function presents precision divergences for probability values smaller than $10^{-15}$. However, it does not presuppose a constraint given that the high-risk values considered for operations are in the order of $10^{-6}$ \citep{uriot2021spacecraft,merz2019risk}.

We also decided to compare the processing time for the presented upper bound expression with the Chan's method, a computationally fast method to calculate the probability of collision \citep{chan1997collision}. Thus, we calculated the processing times of the method proposed by Chan for the probability of collision, using the first approximation order ($M=1$); the upper bound expression presented in~\eqref{eq:upper-bound}; and the upper bound expression presented in~\eqref{eq:derivation-upper-bound-erf} that makes use of the error function.

\begin{figure}[h!]
    \centering
    \includegraphics[scale=0.45]{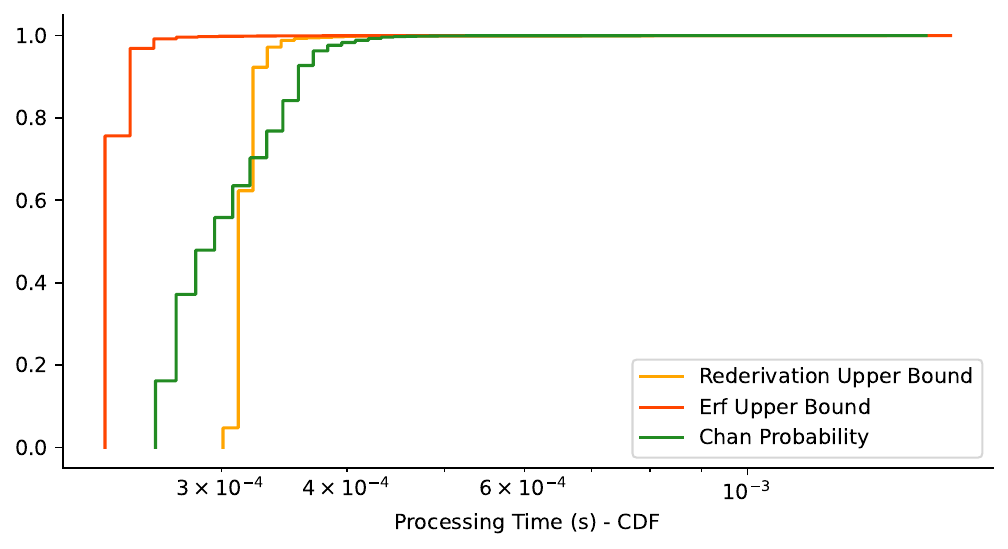}
    \caption{Cumulative distribution function of the processing times (in seconds) of the Chan's method, upper bound using integration and upper bound computation using error function.}
    \label{fig:upper_bound_time_and_chan_cdf}
\end{figure}
As can be seen in Fig.~\ref{fig:upper_bound_time_and_chan_cdf}, if we use the error function to calculate the upper bound, we obtain the shorter processing time. Chan's method has a processing time of $0.31$ ms. When compared to the implementation using the integration we see that they are equivalent in processing time, however when compared to the implementation using the error function, we see that it is approximately $1.3$ times faster than Chan's method.

\section{Conclusion}

We present a derivation from first principles of the method proposed by Akella and Alfriend, a widely used method today for calculating the probability of collision for short-term encounters, presenting a simple and equivalent expression. With the expression obtained, transforming the integration space to a square centered on the origin, it is possible to define easy-to-compute upper and lower bounds as the product of two one-dimensional integrals, which can be computed by making use of the error function. The upper bound can be used to obtain a slightly larger but safe value of the probability of collision, with the advantage of being computationally faster. Our accurate and fast-to-compute bounds can cut the conjunction screening time by eightieth when compared to the method proposed by Akella and Alfriend for the probability of collision, making it possible to evaluate large numbers of conjunctions in close to real-time.

\section{Acknowledgements}

This work is supported by NOVA LINCS (UIDB/04516/2020) with the financial support of FCT.IP. This research was carried out under Project “Artificial Intelligence Fights Space Debris” Nº C626449889-0046305 co-funded by Recovery and Resilience Plan and NextGeneration EU Funds, www.recuperarportugal.gov.pt.

\includegraphics[scale=0.02]{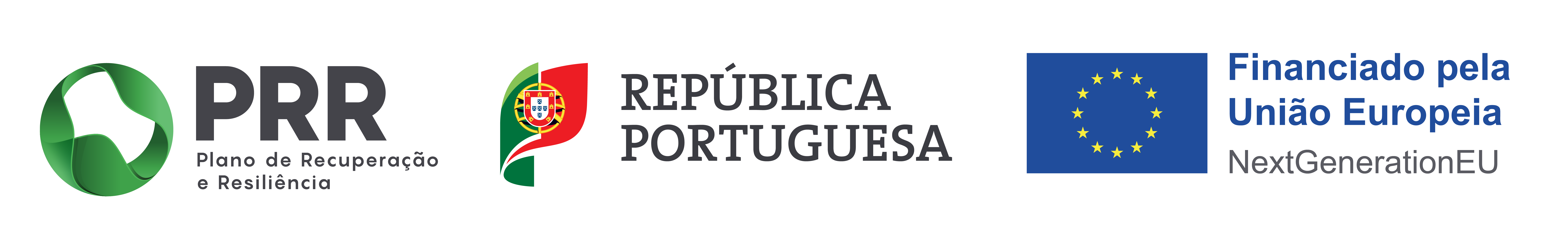}

\bibliographystyle{unsrtnat}
\bibliography{references}

\begin{thebibliography}{29}
\providecommand{\natexlab}[1]{#1}
\providecommand{\url}[1]{\texttt{#1}}
\expandafter\ifx\csname urlstyle\endcsname\relax
  \providecommand{\doi}[1]{doi: #1}\else
  \providecommand{\doi}{doi: \begingroup \urlstyle{rm}\Url}\fi

\bibitem[Kessler and Cour-Palais(1978)]{kessler1978collision}
Donald~J Kessler and Burton~G Cour-Palais.
\newblock Collision frequency of artificial satellites: The creation of a
  debris belt.
\newblock \emph{Journal of Geophysical Research: Space Physics}, 83\penalty0
  (A6):\penalty0 2637--2646, 1978.

\bibitem[Lemmens and Letizia(2022)]{lemmens2020esa}
Stijn Lemmens and Francesca Letizia.
\newblock {ESA}’s annual space environment report.
\newblock Technical report, Technical Report GEN-DB-LOG-00288-OPS-SD, ESA Space
  Debris Office, 2022.

\bibitem[Kleinig et~al.(2022)Kleinig, Smith, and Capon]{kleinig2022collision}
Thomas Kleinig, Brenton Smith, and Christopher Capon.
\newblock Collision avoidance of satellites using ionospheric drag.
\newblock \emph{Acta Astronautica}, 2022.

\bibitem[Bonnal et~al.(2020)Bonnal, McKnight, Phipps, Dupont, Missonnier,
  Lequette, Merle, and Rommelaere]{bonnal2020just}
Christophe Bonnal, Darren McKnight, Claude Phipps, C{\'e}dric Dupont, Sophie
  Missonnier, Laurent Lequette, Matthieu Merle, and Simon Rommelaere.
\newblock Just in time collision avoidance--a review.
\newblock \emph{Acta Astronautica}, 170:\penalty0 637--651, 2020.

\bibitem[Mishne and Edlerman(2017)]{mishne2017collision}
David Mishne and Eviatar Edlerman.
\newblock Collision-avoidance maneuver of satellites using drag and solar
  radiation pressure.
\newblock \emph{Journal of Guidance, Control, and Dynamics}, 40\penalty0
  (5):\penalty0 1191--1205, 2017.

\bibitem[Reiland et~al.(2021)Reiland, Rosengren, Malhotra, and
  Bombardelli]{reiland2021assessing}
Nathan Reiland, Aaron~J Rosengren, Renu Malhotra, and Claudio Bombardelli.
\newblock Assessing and minimizing collisions in satellite mega-constellations.
\newblock \emph{Advances in Space Research}, 67\penalty0 (11):\penalty0
  3755--3774, 2021.

\bibitem[Le~May et~al.(2018)Le~May, Gehly, Carter, and Flegel]{le2018space}
Samantha Le~May, Steve Gehly, BA~Carter, and Sven Flegel.
\newblock Space debris collision probability analysis for proposed global
  broadband constellations.
\newblock \emph{Acta Astronautica}, 151:\penalty0 445--455, 2018.

\bibitem[Lucken and Giolito(2019)]{lucken2019collision}
Romain Lucken and Damien Giolito.
\newblock Collision risk prediction for constellation design.
\newblock \emph{Acta Astronautica}, 161:\penalty0 492--501, 2019.

\bibitem[Balch et~al.(2019)Balch, Martin, and Ferson]{balch2019satellite}
Michael~Scott Balch, Ryan Martin, and Scott Ferson.
\newblock Satellite conjunction analysis and the false confidence theorem.
\newblock \emph{Proceedings of the Royal Society A}, 475\penalty0
  (2227):\penalty0 20180565, 2019.

\bibitem[Poore et~al.(2016)Poore, Aristoff, Horwood, Armellin, Cerven, Cheng,
  Cox, Erwin, Frisbee, Hejduk, et~al.]{poore2016covariance}
Audrey~B Poore, Jeffrey~M Aristoff, Joshua~T Horwood, Roberto Armellin,
  William~T Cerven, Yang Cheng, Christopher~M Cox, Richard~S Erwin, Joseph~H
  Frisbee, Matt~D Hejduk, et~al.
\newblock Covariance and uncertainty realism in space surveillance and
  tracking.
\newblock Technical report, Numerica Corporation Fort Collins United States,
  2016.

\bibitem[Klinkrad(2006)]{klinkrad2006space}
Heiner Klinkrad.
\newblock \emph{Space debris: models and risk analysis}.
\newblock Springer Science \& Business Media, 2006.

\bibitem[Pelton and Ailor(2013)]{pelton2013space}
Joseph~N Pelton and WH~Ailor.
\newblock \emph{Space debris and other threats from outer space}.
\newblock Springer, 2013.

\bibitem[Akella and Alfriend(2000)]{akella2000probability}
Maruthi~R Akella and Kyle~T Alfriend.
\newblock Probability of collision between space objects.
\newblock \emph{Journal of Guidance, Control, and Dynamics}, 23\penalty0
  (5):\penalty0 769--772, 2000.

\bibitem[Alfriend et~al.(1999)Alfriend, Akella, Frisbee, Foster, Lee, and
  Wilkins]{alfriend1999probability}
Kyle~T Alfriend, Maruthi~R Akella, Joseph Frisbee, James~L Foster, Deok-Jin
  Lee, and Matthew Wilkins.
\newblock Probability of collision error analysis.
\newblock \emph{Space Debris}, 1\penalty0 (1):\penalty0 21--35, 1999.

\bibitem[Chan(1997)]{chan1997collision}
Ken Chan.
\newblock Collision probability analysis for earth orbiting satellites.
\newblock \emph{Space cooperation into the 21 st century}, pages 1033--1048,
  1997.

\bibitem[Leclair and Sridharan(2001)]{leclair2001probability}
Raymond~A Leclair and Ramaswamy Sridharan.
\newblock Probability of collision in the geostationary orbit.
\newblock In \emph{Space Debris}, volume 473, pages 463--470, 2001.

\bibitem[Alfano(2005)]{alfano2005numerical}
Salvatore Alfano.
\newblock A numerical implementation of spherical object collision probability.
\newblock \emph{The Journal of the Astronautical Sciences}, 53\penalty0
  (1):\penalty0 103--109, 2005.

\bibitem[Patera(2001)]{patera2001general}
Russell~P Patera.
\newblock General method for calculating satellite collision probability.
\newblock \emph{Journal of Guidance, Control, and Dynamics}, 24\penalty0
  (4):\penalty0 716--722, 2001.

\bibitem[Foster and Estes(1992)]{foster1992parametric}
James~Lee Foster and Herbert~S Estes.
\newblock \emph{A parametric analysis of orbital debris collision probability
  and maneuver rate for space vehicles}.
\newblock NASA, National Aeronautics and Space Administration, Lyndon B.
  Johnson Space Center, 1992.

\bibitem[Berend(1999)]{berend1999estimation}
N~Berend.
\newblock Estimation of the probability of collision between two catalogued
  orbiting objects.
\newblock \emph{Advances in Space Research}, 23\penalty0 (1):\penalty0
  243--247, 1999.

\bibitem[Klinkrad et~al.(2006)Klinkrad, Alarc{\'o}n, and
  S{\'a}nchez]{klinkrad2006operational}
Heiner Klinkrad, J~Alarc{\'o}n, and N~S{\'a}nchez.
\newblock Operational collision avoidance with regard to catalog objects.
\newblock In \emph{Space Debris}, pages 215--240. Springer, 2006.

\bibitem[Coppola(2012)]{coppola2012including}
Vincent~T Coppola.
\newblock Including velocity uncertainty in the probability of collision
  between space objects.
\newblock \emph{Advances in the Astronautical Sciences}, 143, 2012.

\bibitem[Arzelier et~al.(2020)Arzelier, Br{\'e}hard, Joldes, Lasserre, and
  Rondepierre]{arzelier2020rigorous}
Denis Arzelier, Florent Br{\'e}hard, Mioara Joldes, Jean-Bernard Lasserre, and
  Aude Rondepierre.
\newblock {Rigorous derivation of Coppola's formula for the computation of the
  probability of collision between space objects}.
\newblock Research Report Rapport LAAS n{\textdegree} 20008, {LAAS-CNRS},
  January 2020.
\newblock URL \url{https://hal.archives-ouvertes.fr/hal-02444341}.

\bibitem[Boyd and Vandenberghe(2004)]{boyd2004convex}
Stephen~P Boyd and Lieven Vandenberghe.
\newblock \emph{Convex optimization}.
\newblock Cambridge university press, 2004.

\bibitem[Horn and Johnson(2012)]{horn2012matrix}
Roger~A Horn and Charles~R Johnson.
\newblock \emph{Matrix analysis}.
\newblock Cambridge university press, 2012.

\bibitem[Uriot et~al.(2021)Uriot, Izzo, Sim{\~o}es, Abay, Einecke, Rebhan,
  Martinez-Heras, Letizia, Siminski, and Merz]{uriot2021spacecraft}
Thomas Uriot, Dario Izzo, Lu{\'\i}s~F Sim{\~o}es, Rasit Abay, Nils Einecke,
  Sven Rebhan, Jose Martinez-Heras, Francesca Letizia, Jan Siminski, and Klaus
  Merz.
\newblock Spacecraft collision avoidance challenge: design and results of a
  machine learning competition.
\newblock \emph{Astrodynamics}, pages 1--20, 2021.

\bibitem[Metz(2020)]{metzthesis}
Sascha Metz.
\newblock Master thesis: Implementation and comparison of data-based methods
  for collision avoidance in lite operations, 07 2020.

\bibitem[Vallado(2001)]{vallado2001fundamentals}
David~A Vallado.
\newblock \emph{Fundamentals of astrodynamics and applications}, volume~12.
\newblock Springer Science \& Business Media, 2001.

\bibitem[Merz et~al.(2019)Merz, Virgili, Braun, et~al.]{merz2019risk}
Klaus Merz, Benjamin~Bastida Virgili, Vitali Braun, et~al.
\newblock Risk reduction and collision risk thresholds for missions operated at
  esa.
\newblock In \emph{Proceedings of the 27th International Symposium on Space
  Flight Dynamics (ISSFD)}, 2019.

\end{thebibliography}


\end{document}